\documentclass[journal]{IEEEtran}
\usepackage{graphicx}
\usepackage{subcaption} 
\usepackage{epstopdf}
\usepackage{dcolumn}
\usepackage{bm}
\usepackage{color}
\usepackage[mathlines]{lineno}
\usepackage[breaklinks=true,colorlinks=true,linkcolor=blue,urlcolor=blue,citecolor=blue]{hyperref}
\usepackage{ragged2e}
\usepackage{enumitem}
\usepackage{booktabs}
\usepackage{amsthm}
\usepackage{amssymb}
\usepackage{amsmath}
\usepackage{tabularx}
\usepackage{algorithm}
\usepackage{xr}
\usepackage{algpseudocode}

\theoremstyle{definition}

\theoremstyle{remark}

\newcolumntype{Y}{>{\centering\arraybackslash}X}
\ifCLASSINFOpdf
\else
\fi

\usepackage{xcolor}
\newcommand{\zq}[1]{\textcolor{black}{#1}}

\begin{document}

\title{Exploration enhances cooperation in the multi-agent communication system}

\author{Zhao~Song, Chen~Shen, Zhen~Wang,~\IEEEmembership{Fellow,~IEEE}, and The Anh Han
\thanks{Z. Song and T.A. Han are with the School of Computing, Engineering and Digital Technologies, Teesside University,  Middlesbrough, UK. (email: z.song@tees.ac.uk; t.han@tees.ac.uk).}
\thanks{C. Shen is with the Faculty of Engineering Sciences, Kyushu University, Fukuoka 816-8580, Japan. (email: steven\_shen91@hotmail.com).}
\thanks{Z. Wang is with the School of Cybersecurity, and School of Artificial Intelligence, OPtics and ElectroNics (iOPEN), Northwestern Polytechnical University, China (e-mail: w-zhen@nwpu.edu.cn).}
}




\maketitle

\begin{abstract}
Designing protocols enhancing cooperation for multi-agent systems remains a grand challenge. Cheap talk, defined as costless, non-binding communication before formal action, serves as a pivotal solution. However, existing theoretical frameworks often exclude random exploration, or noise, for analytical tractability, leaving its functional impact on system performance largely unexplored. To bridge this gap, we propose a two-stage evolutionary game-theoretical model, integrating signalling with a donation game, with exploration explicitly incorporated into the decision-making. Our agent-based simulations across topologies reveal a universal optimal exploration rate that maximises system-wide cooperation. Mechanistically, moderate exploration undermines the stability of defection and catalyses the self-organised cooperative alliances, facilitating their cyclic success. 
Moreover, the cooperation peak is enabled by the delicate balance between oscillation period and amplification.
Our findings suggest that rather than pursuing deterministic rigidity, embracing strategic exploration, as a form of engineered randomness, is essential to sustain cooperation and realise optimal performance in communication-based intelligent systems.
\end{abstract}


\IEEEpeerreviewmaketitle

\section{Introduction}
Achieving coordination among autonomous agents remains a cornerstone challenge in the fields of multi-agent systems (MAS)~\cite{tan2016evolutionary,sun2022stimulating,zhou2022swarm,guo2023third,HAN202633}. In many decentralised environments, self-interested agents often face a conflict between maximising local utility and achieving system-level optimality, a scenario mathematically formalised as a social dilemma. Without effective intervention mechanisms, such systems are prone to collapsing into ``Nash Equilibria" of non-cooperation, leading to suboptimal collective performance~\cite{nash2024non,fang2020distributed}. Over the past decades, researchers have proposed various structural and algorithmic solutions to resolve these deadlocks, ranging from reputation~\cite{fu2008reputation,brandt2003punishment,chen2016evolution}, tag-mediated~\cite{hadzibeganovic2019nonconformity,masuda2007tag,riolo2001evolution}, and incentive mechanisms~\cite{szolnoki2010reward,cimpeanu2021cost,han2022institutional,semonsen2018opinion,HAN202633}, for sustainable cooperative outcomes. In addition, the hybrid system consisting of humans and intelligent machines attracts increasing attention to ensure cooperative relationships~\cite{paiva2018engineering,andras2018trusting,guo2023facilitating,zimmaro2024emergence,an2023cooperative}.


Among these solutions, cheap talk, functioning as a costless, non-binding signalling protocol, has emerged as a pivotal mechanism for enabling adaptive coordination~\cite{farrell1996cheap,skyrms2010signals,crawford1982strategic}. Unlike costly enforcement mechanisms, cheap talk facilitates ``intention sharing" prior to action, serving as a lightweight information exchange channel that incurs negligible computational or transmission overhead~\cite{santos2011co,han2015synergy,robson1990efficiency}. In engineering applications, such protocols are ubiquitous for resolving resource conflicts and synchronising behaviours, exemplified by tasks such as negotiation in automated trading markets and collision avoidance signalling in autonomous vehicle swarms~\cite{wellman2001auction,wooldridge2009introduction}. In addition, both theoretical models and agent-based simulations consistently demonstrate that implementing such pre-play communication layers can significantly boost the emergence of cooperative behaviours, transforming the system's payoff landscape without altering the underlying physical constraints~\cite{krakauer1995spatial,torney2011signalling,wiseman2001cooperation}.


However, most existing theoretical frameworks analysing cheap talk are based on the assumption of rare exploration for analytical tractability~\cite{wu2012small,hindersin2019computation,song2026network}.
Exploration, in this context, refers to the stochastic departure from the prescribed or optimal behaviour pattern.
In contrast, stochasticity is intrinsic to real-world systems, manifesting variously as genetic mutations in evolutionary processes, or error in multi-agent learning systems~\cite{houser2002revisiting,grujic2014comparative,de2021greed}. On the one hand, theoretical studies consistently show that such non-zero levels of exploration are required to prevent evolutionary systems from being trapped in absorbing states or can significantly alter equilibrium outcomes~\cite{chen2024number,mobilia2007phase,traulsen2010human}. 
On the other hand, the introduction of cheap talk inherently expands the strategic landscape, making strategic uncertainty a pervasive phenomenon. Consequently, a critical scientific question arises: \textit{how does cheap talk sustain cooperation in the presence of random exploration, and specifically, how does such stochasticity influence the evolutionary success of diverse cooperative strategies?}

To address this critical gap, we develop a theoretical model under the framework of evolutionary game that incorporates exploration into the decision-making process of a two-stage cheap talk game. Specifically, our model integrates a pre-game communication stage, where players decide whether to send a cooperative signal, with a subsequent in-game decision-making stage, where players engage in a Donation Game \cite{sigmund2010calculus}. We analyse the complete strategy space of eight possible behaviours, categorising them into intuitive strategies (simple consistency) and deliberative strategies (conditional reasoning) to reflect varying cognitive processes. Furthermore, we introduce exploration, or behavioural error, during population updates, where players may randomly adopt arbitrary strategies rather than imitating successful role models. 
It is important to clarify that in this study, the term exploration is distinct from perception noise or response noise, typically in signal transmission or decision-making.
Our model builds upon the framework established in \cite{song2026network}, where a minimal exploration is used to mitigate the finite-size effect; the present work elevates exploration to a central strategic variable, focusing on the implications of the exploration process itself.

Based on this theoretical model, we conduct agent-based simulations in structured populations, focusing on the emergence of cooperation in noisy cheap talk scenarios. First, we confirm that even with stochastic interference, cheap talk combined with network reciprocity remains a robust pathway for sustaining cooperation. Crucially, we discover the existence of an optimal, non-zero strategic exploration rate that maximises the system's overall level of cooperation. This optimal rate facilitates the stable success of both intuitive and deliberative cooperative strategies. Mechanistically, moderate exploration functions by destabilising entrenched defection clusters while enabling cooperative strategies to organise into compact, resilient spatial clusters. Finally, we note that while exploration serves as a powerful optimisation force, it is not a panacea; under strict conditions, such as extreme dilemma intensity or scenarios where cheap talk loses utility, the benefits of the optimal exploration rate diminish.

Our main contributions are summarized as follows:
\begin{itemize}
    \item  By relaxing the conventional assumption of rare exploration, we introduce a two-stage evolutionary game model in which a controlled variant characterising intrinsic stochasticity is incorporated. This enables a rigorous quantification of the transition from deterministic rigidity to stochastic coordination.
    \item We provide theoretical evidence for the robust persistence of cooperation in multi-agent communication systems under non-zero exploration rates. This result bridges the gap between idealised noise-free models and real-world complex systems where uncertainty is inherent.
    \item We uncover a cyclic success mechanism of cooperation. Unlike static equilibria in the absence of exploration, this mechanism, driven by exploration, ensures the resilience of cooperation through a periodic self-organisation process. We show that the formation of compact cooperative clusters effectively disrupts the dominance of defectors, preventing the system from collapsing into non-cooperative traps.
    \item We identify an optimal exploration rate that maximises system-wide cooperation, enabled by the delicate balance between periodic oscillations and amplification. Furthermore, our findings emphasise the essential power of variety, highlighting the controlled stochasticity rather than deterministic stability for global coordination.
\end{itemize}

The remainder of this article is organized as follows. In Section \ref{Preliminaries}, we give the preliminaries. Section \ref{Model and Method} presents the model and method. In Section \ref{Results}, we show the agent-based simulation results of how exploration influences. In Sections \ref{Discussion} and \ref{Conclusion}, we discuss and conclude the results of this article. 

\section{Preliminaries}
\label{Preliminaries}
Social dilemmas typically arise from the fundamental conflict between collective interests and individual motivations. To quantitatively analyse this conflict, researchers frequently employ the two-player two-strategy game as the foundational theoretical model. Within this framework, two participants simultaneously choose to either cooperate ($C$) or defect ($D$). The outcomes of their interaction are defined by four standard payoff parameters: a reward $R$ for mutual cooperation, a punishment $P$ for mutual defection, a temptation $T$ for a defector exploiting a cooperator, and a sucker's payoff $S$ for a cooperator exploited by a defector. 

Based on the relative ranking of $T, R, P,$ and $S$, this framework encompasses four fundamental game models, each characterized by distinct Nash Equilibria (NE)~\cite{yang2013towards}.
In the Prisoner’s Dilemma game (PD) with $T > R > P > S$, defection is the strictly dominant strategy, leading to a unique, Pareto-inefficient NE at $(D, D)$. Conversely, the Snowdrift game (SDG) with $T > R > S > P$ lacks a dominant strategy, typically resulting in a mixed-strategy NE or stable polymorphism. The Stag Hunt game (SHG) with $R > T > P > S$ presents a coordination dilemma characterized by two pure-strategy NE: a payoff-dominant $(C, C)$ and a risk-dominant $(D, D)$. Finally, the Harmony game (HG) with $R > T > S > P$ facilitates seamless coordination, where the unique NE naturally converges to mutual cooperation $(C, C)$.

Among these models, the Donation Game is widely regarded as a standard paradigm for studying the evolution of cooperation, essentially serving as a symmetric and simplified version of the Prisoner's Dilemma. In the Donation Game, a cooperator incurs a cost $c$ to provide a benefit $b$ to the opponent (where $b > c$), while a defector pays and provides nothing. This simplifies the payoff matrix to $R = b-c$, $T = b$, $S = -c$, and $P = 0$. To further generalise the analysis, the dilemma strength $r = c/(b-c)$ is introduced. This parameter effectively characterises the difficulty of cooperation: a large $r$ denotes a strong dilemma strength, where evolutionary pressure suppresses the survival of cooperation, while a small $r$ represents a weak dilemma strength, allowing for its emergence~\cite{axelrod1988further,wang2015universal}.

Denote $\mathcal{G=\{\mathcal{V},\mathcal{E}}\}$ as the topology structure, where $\mathcal{V}=\{1,2,\dots,\mathcal{N}\}$ refers to the node set, and $\mathcal{E}$ refers to the link set. Denote $a_{ij}\in \mathbb{R}$ as the element of the adjacency matrix, where $a_{ij}=1$ indicates player $i$ has a connection with player $j$, whereas $a_{ij}=0$ indicates no connection. Here, we consider undirected networks, and the degree of each node is $k_i=\sum^N_{j=1}a_{ij}$. $\mathcal{G}$ is a homogeneous network if $k_i=k_j, \forall i,j \in \mathcal{V}$, otherwise heterogeneous. Specifically, when $k_i=N-1, \forall i \in \mathcal{V}$, $\mathcal{G}$ is a well-mixed network, or complete graph.
\begin{table*}[t]
\centering
\caption{The strategy space defined by the triplet $(u, p, q)$. $u=1$ denotes active signalling, $u=0$ denotes silence. $p$ and $q$ denote the action in a game, responding to a signal or silence, respectively.
}
\begin{tabular*}{0.9\textwidth}{@{\extracolsep{\fill}}l|c|c|c|l|l}
\hline
\textbf{Strategy} & \textbf{Signal} ($u$) & \textbf{Resp. to Signal} ($p$) & \textbf{Resp. to Silence} ($q$) & \textbf{Strategy Description} &\textbf{Strategy Type} \\
\hline
\textbf{ACC} & 1 (A) & 1 (C) & 1 (C) & Unconditional Cooperation & Intuitive strategy \\
\textbf{ACD} & 1 (A) & 1 (C) & 0 (D) & Conditional Cooperation & Deliberative strategy\\
\textbf{ADC} & 1 (A) & 0 (D) & 1 (C) & Paradoxical Cooperation & Deliberative strategy\\
\textbf{ADD} & 1 (A) & 0 (D) & 0 (D) & Strategic Defection & Deliberative strategy\\
\hline
\textbf{NCC} & 0 (N) & 1 (C) & 1 (C) & Silent Cooperation & Deliberative strategy\\
\textbf{NCD} & 0 (N) & 1 (C) & 0 (D) & Conditional Defection & Deliberative strategy\\
\textbf{NDC} & 0 (N) & 0 (D) & 1 (C) & Specious Cooperation & Deliberative strategy\\
\textbf{NDD} & 0 (N) & 0 (D) & 0 (D) & Unconditional Defection & Intuitive strategy \\
\hline
\end{tabular*}
\label{tab:strategies}
\end{table*}
\section{Model and Method}
\label{Model and Method}
\subsection{The Two-Stage Cheap Talk Game}
\begin{figure*}[htb]
    \centering
    \includegraphics[width=\linewidth]{ 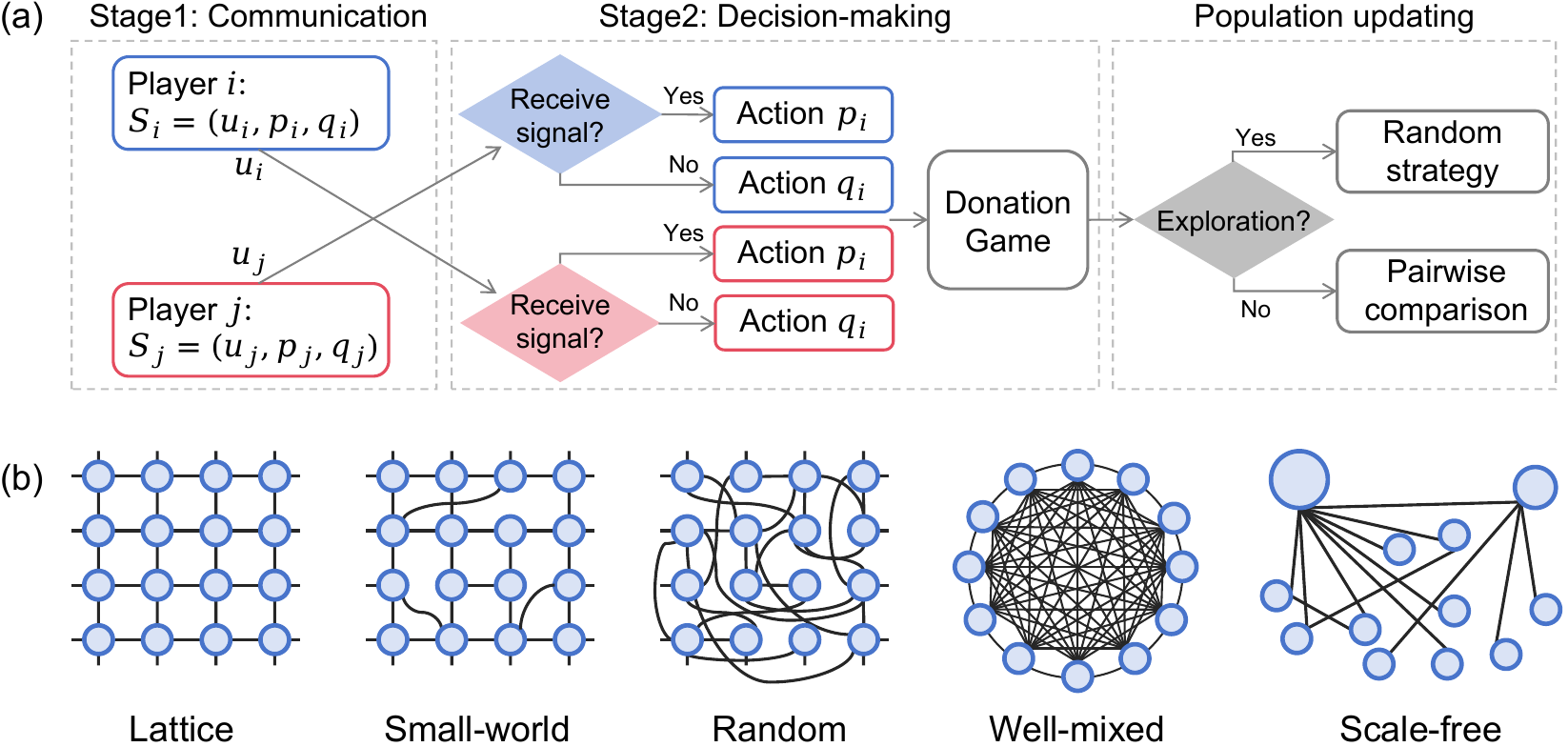}
    \caption{\textbf{Two-stage cheap talk game and population structures.}
    (a) The game includes two stage: pre-game communication and in-game decision-making. In each evolutionary update step, a focal player, either explores a new random strategy or learn the strategy from a random role model.
    (b) Examples of the investigated population structures: square lattice, small-world, random, well-mixed, and scale-free networks.
    }
    \label{fig:model}
\end{figure*}

We model the interaction as a two-stage evolutionary game involving pre-game communication and in-game decision-making, as shown in Figure \ref{fig:model}(a). In the first stage, players choose whether to send a cooperative signal ($A$) or remain silent ($N$). This signal is cheap talk: non-binding and carries no intrinsic transmission cost. In the second stage, players engage in a Donation Game, simultaneously choosing to either cooperate ($C$) or defect ($D$) based on the signals exchanged in the first stage.



Formally, player $i$'s strategy can be defined by a
triplet $S_i = (u_i, p_i, q_i) \in \{0,1\}^3$, where $u_i$ is the signal ($1$ for $A$, $0$ for $N$), $p_i$ is the action if the co-player signals $A$ ($1$ for $C$, $0$ for $D$), and $q_i$ is action if the co-player is silent $N$ ($1$ for $C$, $0$ for $D$).
This formulation gives rise to $2^3 = 8$ distinct strategies, listed in Table~\ref{tab:strategies}.

We categorise these strategies into two types based on dual process theory to reflect reasoning complexity:
\begin{itemize}
\item \textbf{Intuitive strategies} ($u_i=p_i=q_i$): These players exhibit consistent, heuristic-based behaviours where the pre-play signal aligns with subsequent actions regardless of the opponent's signal. This category includes unconditional cooperation (ACC) and unconditional defection (NDD). We assume such consistency incurs no extra reasoning cost.
\item \textbf{Deliberative strategies} ($u_i \neq p_i$ or $p_i \neq q_i$): These players engage in complex reasoning processes, such as conditional behaviours (e.g., ACD, responding differently to distinct signals) or deceptive signalling (e.g., ADD or NDC, where the signal contradicts the action). To model the effort required for such decision-making processes, we impose a reasoning cost $\gamma \in [0, 1]$. 
\end{itemize}
Crucially, this cost applies if and only if the player adopts a deliberative strategy. The cost function $C(S_i)$ is defined as:
\begin{equation}
    C(S_i)=\left\{
    \begin{array}{ll}
        0, & \text{if } u_i=p_i=q_i \\
        1, & \text{otherwise.}
    \end{array}
    \right.
\end{equation}

\subsection{Payoff Formulation}
\zq{Suppose players $i$ and $j$ are paired in a game.} Let $x$ be the probability that player $i$ cooperates, and $y$ be the probability that player $j$ cooperates. In the deterministic case, these probabilities depend on the opponent's signal:
\begin{equation}
    x = u_j p_i + (1-u_j) q_i, \quad y = u_i p_j + (1-u_i) q_j.
\end{equation}
The expected payoff for player $i$ against player $j$ is given by the bilinear form:
\begin{equation}
    \Pi(S_i, S_j) = \begin{pmatrix} x & 1-x \end{pmatrix} \mathbf{G} \begin{pmatrix} y \\ 1-y \end{pmatrix}^T - \gamma C(S_i).
    \label{payoff}
\end{equation}

\subsection{Population Updates}
\label{subsection:pop_structure}
We analyse the evolutionary dynamics within a spatially structured population, which provides the environment for local interactions. To ensure the robustness of our findings, the population is modelled using three important types of regular networks: lattice, small-world, and random networks \cite{szabo2007evolutionary}. In these homogeneous topologies, players have on average $k=4$ neighbours.
Additionally, we consider well-mixed networks that do not provide network reciprocity, as well as scale-free networks characterised by heterogeneity, as illustrated in Figure \ref{fig:model}(b).

Each player $i$ occupies a fixed node in $\mathcal{V}$ and interacts exclusively with their set of first-order neighbours $\Omega_i$, via links $\mathcal{E}$. Initially, each node is populated with a strategy chosen uniformly at random from the eight available types. In every time step, each player $i$ accumulates a total payoff, $\phi_i$, by playing the game against all neighbours: $$\phi_i = \sum_{j \in \Omega_i} \Pi(S_i, S_j),$$ where $\Pi(S_i, S_j)$ corresponds to the expected pairwise payoff derived in Equation (\ref{payoff}).

\subsection{Exploration and Imitation}
The evolution of strategies is governed by social learning incorporating random exploration. We model the population dynamics using an asynchronous Monte Carlo simulation on spatially structured populations. In each time step, a focal player $i$ is chosen at random and updates their strategy via one of two mechanisms. 
With a small probability $\mu$, the player adopts a strategy selected uniformly at random from the eight available types. This mechanism corresponds to \textit{strategic exploration} (or the $\epsilon$-greedy strategy in reinforcement learning). It is important to distinguish this exploration from perception noise (errors in signal transmission); here, we focus strictly on the former to investigate the role of strategy exploration.
With probability $1-\mu$, player $i$ attempts to learn from a neighbour. A role model $j$ is selected randomly from $i$'s neighbours $\Omega_i$. Player $i$ adopts $j$'s strategy with a probability determined by the Fermi function~\cite{sigmund2010social}:
\begin{equation}
    W(S_j \to S_i) = \frac{1}{1+e^{\beta(\phi_i-\phi_j)}},
\end{equation}
where $\phi_i$ and $\phi_j$ are the total accumulated payoffs of player $i$ and $j$, respectively. The parameter $\beta$ represents the intensity of selection; a higher $\beta$ implies that players are more likely to imitate neighbours with strictly higher payoffs.

\begin{algorithm}[tb]
\centering
    \caption{Agent-based simulation}
    \label{simu-1}
    \algnewcommand\Input{\item[\textbf{Input:}]}
    \algnewcommand\Output{\item[\textbf{Output:}]}
    \begin{algorithmic}[1]
        \State \textbf{Initialization:} 
        Arrange players on nodes, and assign each player a random strategy.
        
        \For{$t = 1$ to $T_{max}$}
            \For{$i = 1$ to $\mathcal{N}$} \Comment{One Monte Carlo Step (MCS)}
                \State Randomly select a focal player $i$ from $\mathcal{V}$;
                \State Calculate total payoff $\phi_i$ by interacting with  $\Omega_i$;
                \State Randomly select a neighbor $j$ from $\Omega_i$;
                \State Calculate total payoff $\phi_j$ by interacting with $\Omega_j$;
                
                \State Generate a random number $r_1 \in [0, 1]$;
                \If{$r_1 < \mu$} \Comment{Exploration mechanism}
                    \State $s_i \leftarrow$ Random strategy from strategy set
                \Else \Comment{Imitation mechanism}
                    \State Calculate Fermi probability:
                    \State $W \leftarrow \frac{1}{1+\exp\left[ \beta(\phi_i - \phi_j)\right]}$;
                    \State Generate a random number $r_2 \in [0, 1]$;
                    \If{$r_2 < W$}
                        \State $s_i \leftarrow s_j$;
                    \EndIf
                \EndIf
            \EndFor
            
            \If{$t > T_{max} - T_{avg}$}
                \State Record current fraction of cooperators $\rho_C(t)$;
            \EndIf
        \EndFor
    \end{algorithmic}
\end{algorithm}

\section{Experimental Results}
\label{Results}
\subsection{Implementation Details}
The asynchronous Monte Carlo simulation proceeds using the update protocol described in the above Section (see Algorithm \ref{simu-1}). One Monte Carlo Step (MCS) consists of $\mathcal{N}$ 
such elementary updates, ensuring that each player is chosen to update their strategy once on average per MCS. For our simulations, we set the population size to $\mathcal{N}=50^2$. The selection intensity is set to $\beta=10$ to model strong selection. Simulations are run for $3\times 10^4$ MCS, and data are averaged over the final 5,000 steps to ensure the system has reached a statistical steady state.

\subsection{Cheap talk remains a powerful tool for cooperation}
\begin{figure*}[h]
    \centering
    \includegraphics[width=\linewidth]{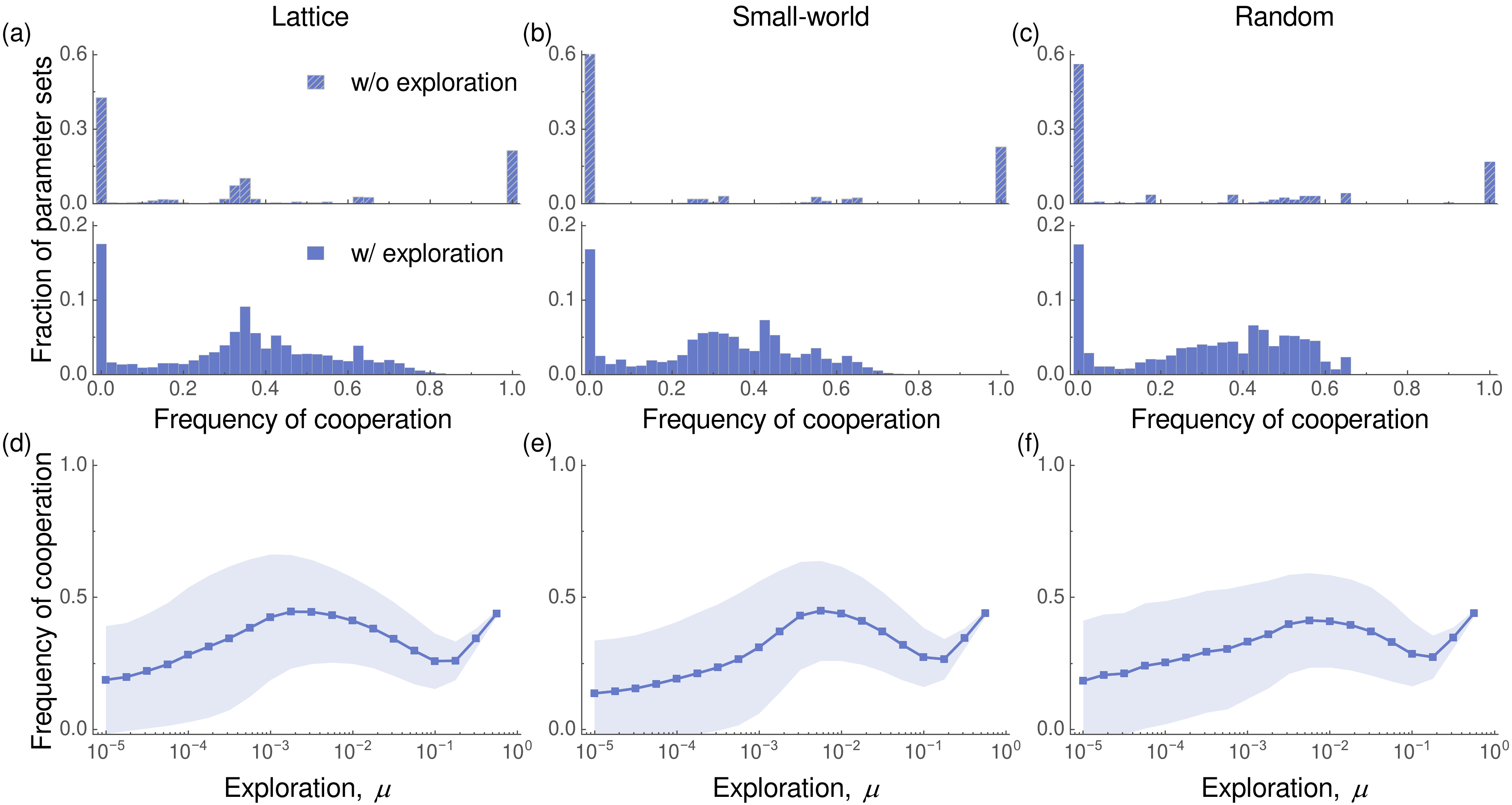}
    \caption{\textbf{Cheap talk sustains cooperation in the exploration scenarios, with optimal exploration maximising the overall cooperation. }
    Panels (a)-(c) show the frequency of cooperation across parameter sets without and with exploration; panels (d)-(f) show the average frequency and the standard deviation of cooperation as a function of exploration. 
    Shown are the results of independent simulations on lattice, small world, and random networks, respectively and uniformly sampled $r \in [0,0.3]$, $\gamma \in [0,0.3]$, and $\mu \in [10^{-5},10^{-\frac{3}{4}}]$.
    }
    \label{fig1}
\end{figure*}
\begin{figure*}[h]
    \centering
    \includegraphics[width=\linewidth]{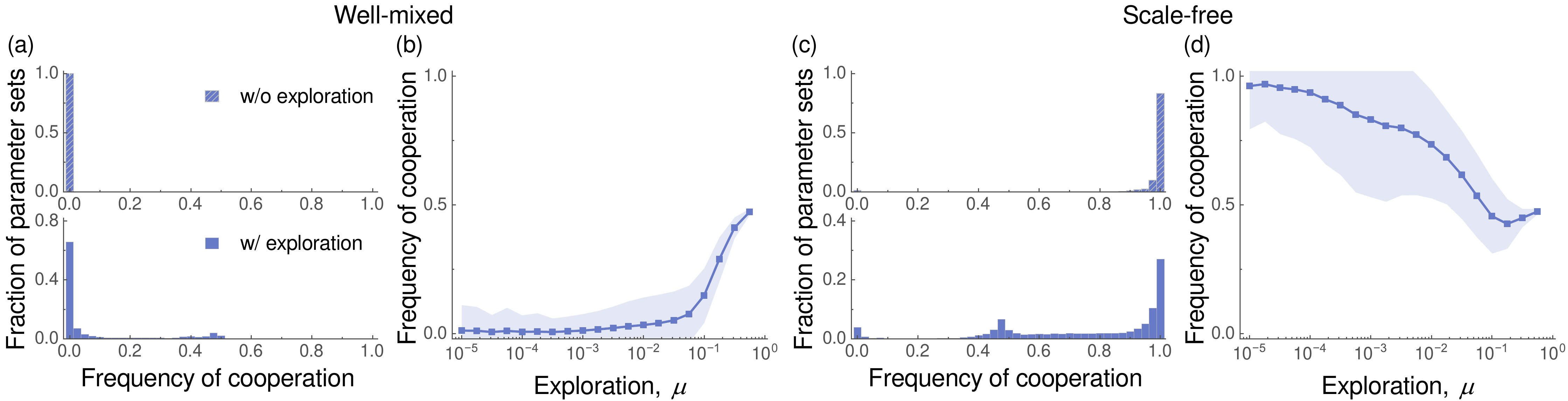}
    \caption{\textbf{Cheap talk can not sustain cooperation in the absence of network reciprocity, while exploration disturbs cooperation on heterogeneous networks. }
    Panels (a) and (c) show the frequency of cooperation across parameter sets without and with exploration; panels (b) and (d) show the average frequency and the standard deviation of cooperation as a function of exploration. 
    Shown are the results of independent simulations on well-mixed and scale-free networks, respectively and uniformly sampled $r \in [0,0.3]$, $\gamma \in [0,0.3]$, and $\mu \in [10^{-5},10^{-\frac{3}{4}}]$.
    }
    \label{fig7}
\end{figure*}

\begin{figure*}[h]
    \centering
    \includegraphics[width=\linewidth]{ 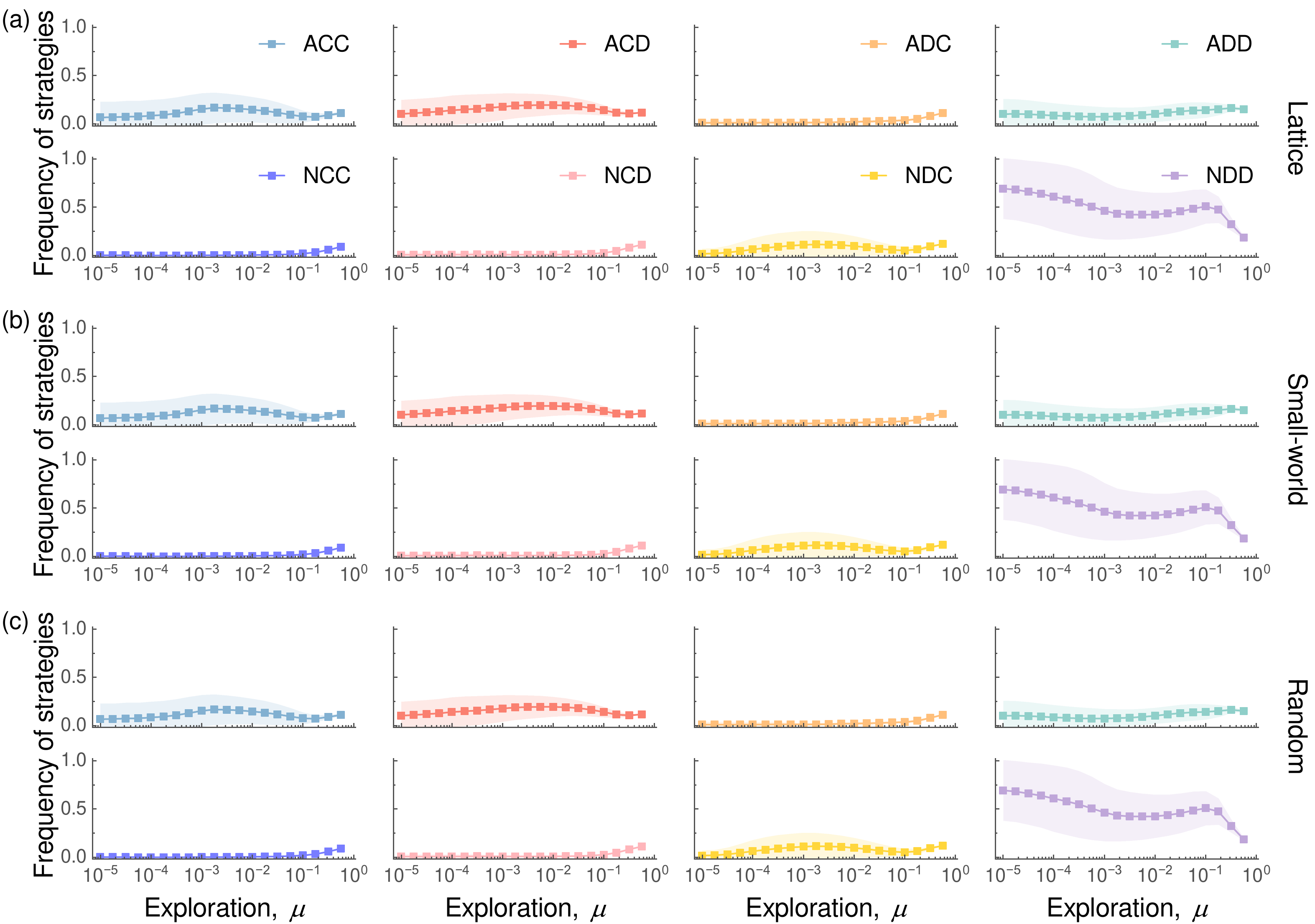}
    \caption{\textbf{exploration enables the optimal success of cooperative strategies across both intuitive and deliberative types. }
    Panels show the average frequency and the standard deviation of each strategy as a function of exploration on lattice, small world, and random networks, from top to bottom. 
    Shown are the results of 100,000 simulations and randomly sampling from uniform distributions of $\mu \in [10^{-5},10^{\frac{3}{4}}]$, $r \in [0,0.3]$, $\gamma \in [0,0.3]$.}
    \label{fig2}
\end{figure*}

Our results demonstrate that, together with network reciprocity, cheap talk remains a powerful tool for sustaining cooperation in the presence of strategic exploration.
We extend our analysis beyond regular lattices to include small-world and random networks, comparing the distribution of average cooperation frequencies in the absence versus the presence of exploration (top row of Figure \ref{fig1}). In the exploration-free scenario, the distributions across all three topologies are polarised, clustering predominantly at either full defection or full cooperation. This binary behaviour indicates that finite-size effects and absorbing states are universal challenges in finite populations, irrespective of network structure. In contrast, when strategic exploration is introduced, the distributions shift significantly across all topologies: while still skewed, they exhibit a broad, dynamic spread with a significant fraction of outcomes sustaining intermediate to high levels of cooperation
Specifically, the frequency of cooperation from 0 to 0.8 significantly increased, while the frequency of cooperation at 0 is largely reduced, across topologies.
This consistency confirms that strategic exploration serves as a generic mechanism to prevent systems from locking into static defection, allowing cheap talk to robustly sustain cooperation in diverse structural environments.

\subsection{Optimal exploration enhances overall cooperation}
Building on this observation, we move from a qualitative distribution analysis to a quantitative investigation of how exploration governs the cooperation outcomes.

Critically, the system exhibits the optimal exploration rate that maximises cooperative outcomes. As shown in the bottom row of Figure \ref{fig1}, the relationship between average cooperation frequency and exploration rate follows a distinct non-monotonic, inverted U-shape trajectory across lattice, small-world, and regular random networks. Cooperation starts at a baseline level for the lowest exploration rates ($\mu=10^{-5}$), increases gradually, and reaches a clear maximum around $\mu = 10^{-3}$. Beyond this optimum, cooperation declines first and then increases to near 0.5 as the system approaches random behaviour. Although the peak amplitude varies slightly due to the topological properties (e.g., shorter average path lengths in random networks), the existence of this optimal exploration window is robust. This universal pattern indicates that a moderate level of noise, inducing exploration, is essential for maximising the system's cooperative capacity.

To further delineate how strategic exploration impacts the role of cheap talk, we evaluated the system on well-mixed and scale-free networks, revealing the functional boundaries of the above findings. First, in a well-mixed population (Figure \ref{fig7}(a-b)), cooperation fails completely in the absence of exploration and rises only trivially with noise. This control experiment confirms that network reciprocity is the fundamental prerequisite; exploration acts as a catalyst only when spatial structure already enables local clustering. Second, on scale-free networks (Figure \ref{fig7}(c-d)), together with intrinsically strong network reciprocity, cheap talk secures high cooperation. In these highly heterogeneous topologies, the presence of highly connected hubs naturally provides robust structural reciprocity, securing near-perfect cooperation. Consequently, introducing random exploration serves not as a catalyst but as a disruptive force; it likely undermines the dominance of these cooperative hubs, thereby dismantling the structural cohesion required for cooperation and leading to a monotonic decline in performance.
In summary, the optimal exploration effect is specific to distributed, homogenous topologies, including lattice, small world, and random. In these systems, which lack the overpowering hubs of scale-free networks but possess the local structure of lattices, an intermediate exploration rate is critical to break local deadlocks and unleash the full potential of cheap talk.


\subsection{Exploration enables the success of cooperative strategies}
To uncover the mechanistic basis of the observed cooperation peaks, we transition from aggregate outcomes to a detailed analysis of the strategy compositions.

The enhancement of cooperation is driven by the selective promotion of cooperative strategies. Figure \ref{fig2} details the frequency of each strategy as a function of exploration rate across the three network topologies, and detailed distribution can be found in Figures 
S1, S4 and S7
. Consistent with the aggregate cooperation peak in Figure \ref{fig1}, the optimal exploration rate (around $\mu = 10^{-3}$) facilitates the simultaneous success of three key cooperative strategies: the intuitive unconditional cooperation (ACC), the deliberative conditional cooperation (ACD), and the parasitic specious cooperation (NDC). In all network types, these strategies exhibit a coordinated peak around the optimal rate, while the dominant unconditional defection (NDD) significantly declines. This differential impact confirms that moderate strategic exploration boosts cooperation by specifically leveraging the synergy between signalling and conditional responses, rather than merely introducing randomness. 

\begin{figure*}[tb]
    \centering
    \includegraphics[width=1\linewidth]{ 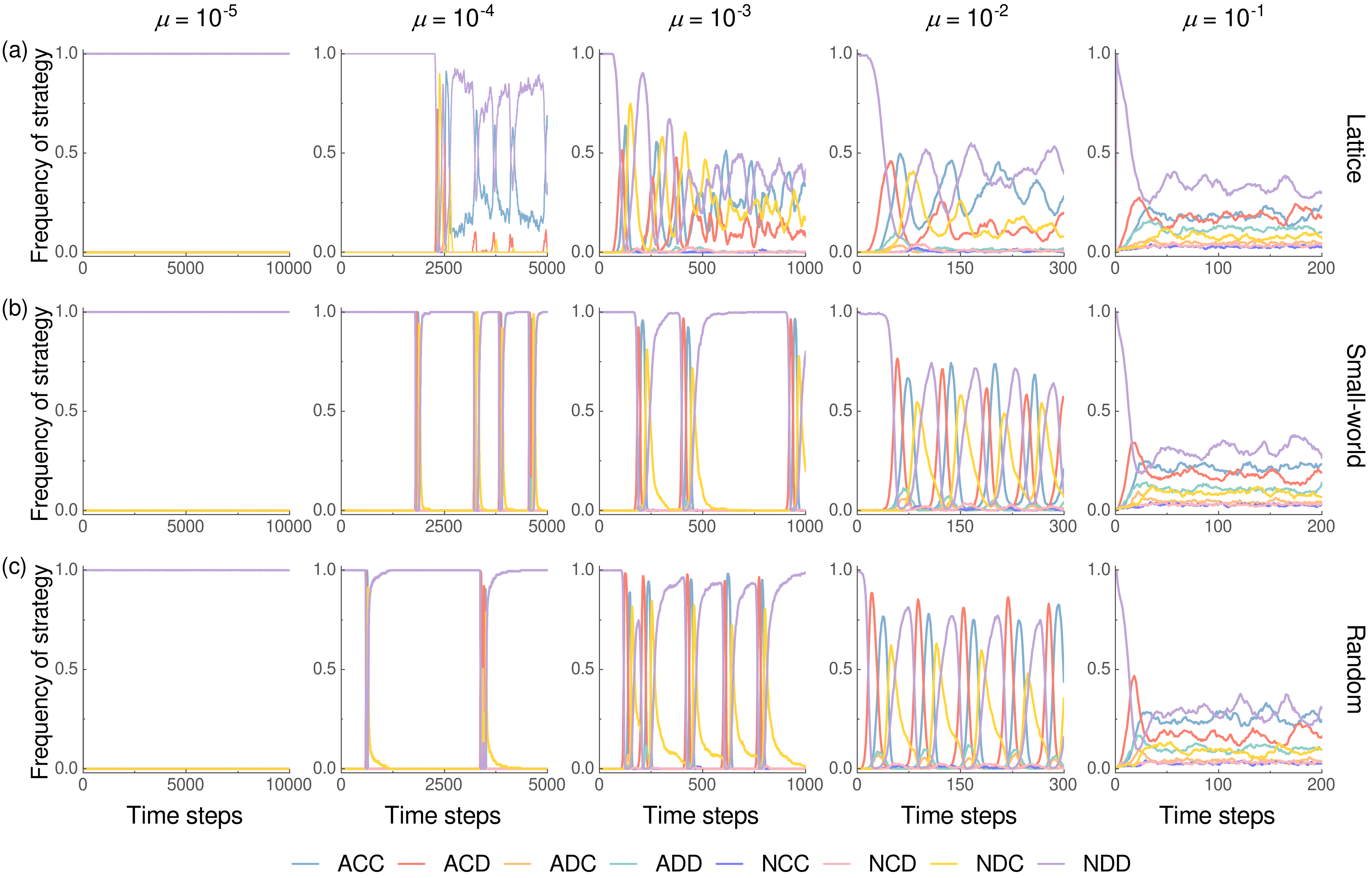}
    \caption{\textbf{Moderate exploration maximises cooperation by balancing the amplitude and period of cyclic success.}
    Shown is the time evolution of strategy frequencies starting from a population of full unconditional defection (NDD). 
    exploration lets the phase lag between NDD (purple) $\rightarrow$ ACD (red)$\rightarrow$ ACC (blue)$\rightarrow$ NDC (yellow).
    Parameters are set as $r=0.02$, $\gamma=0.1$, $\mu=10^{-5}, 10^{-4}, 10^{-3}, 10^{-2}$, and $10^{-1}$ from the left column to the right column, respectively.
    }
    \label{fig5}
\end{figure*}

\subsection{Optimal exploration balances the oscillation period and amplification.}
To reveal the dynamic nature of this cooperation, we investigate the time evolution of strategy frequencies, identifying a cyclic succession driven by the optimal exploration. Figure \ref{fig5} presents the evolutionary dynamics starting from a full defection state across the three network topologies with various exploration rates.

Moderate exploration rates, balances between the period and amplitude of cyclic success, thereby maximising the overall cooperation level. At the very low exploration rate ($\mu=10^{-5}$, first column), the system in all networks remains trapped in the initial absorbing state of NDD dominance due to insufficient exploration. 
As the exploration rate increases ($\mu=10^{-4}$, second column), the system is characterised by long-period, high-amplitude oscillations. In this regime, exploration is rare. When a cooperative mutant (ACD) finally invades the defector-dominated population, it triggers a system-wide boom of cooperation due to the slow timescale of subsequent invasions. However, this is followed by a dramatic collapse as specious cooperation slowly accumulates and eventually destroys the overall cooperation. These cycles exhibit a distinct phase lag between NDD $\to$ ACD $\to$ ACC $\to$ NDC $\to$ NDD. While the peak cooperation is high, the long inter-arrival times between waves result in prolonged periods of NDD dominance, lowering the long-term averaged cooperation. Conversely, at high exploration rates ($\mu=10^{-2}, 10^{-1}$, the right two columns), the system shifts to short-period, low-amplitude fluctuations. Excessive exploration acts as a disruptive noise, inducing new strategies appear everywhere simultaneously and therefore, desynchronizing the population. Consequently, the system settles into a noisy quasi-steady state, where strategy frequencies vibrate rapidly within a narrow range. Although cooperation is continuous, it is suppressed at a mediocre level because cooperation is fragmented by noise before it can grow to dominance. Crucially, at a moderate exploration rate ($\mu=10^{-3}$, middle column), strikes a critical balance between oscillation amplitude and period. At this optimal exploration, exploration is frequent enough to break the NDD deadlocks quickly, shortening the low-cooperation periods, but moderate enough to preserve the spatial integrity of cooperation, maintaining high-cooperation amplitudes. 

\begin{figure}[tb]
    \centering
    \includegraphics[width=\linewidth]{ 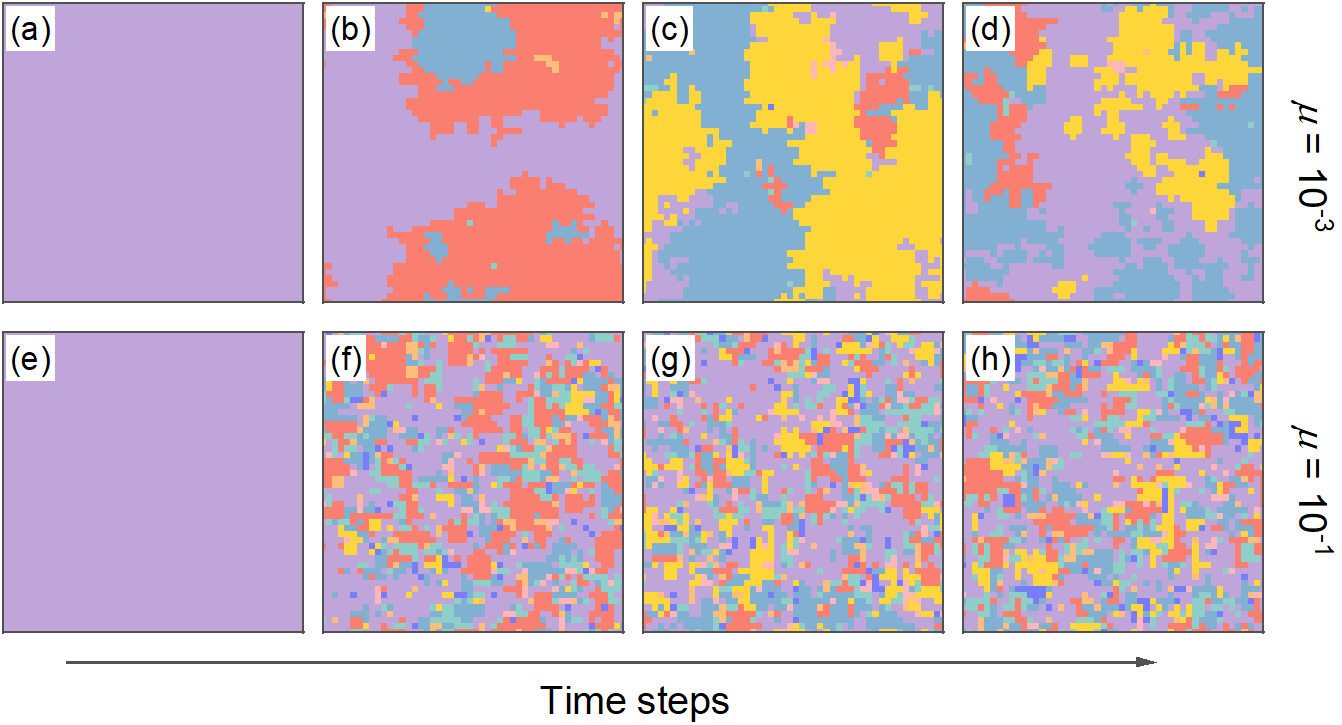}
    \caption{\textbf{Moderate exploration enables the cooperation alliance of intuitive and deliberative strategies.}
    Panels show the strategy distribution during the evolution process on lattice network. 
    Parameters are set as: $r=0.02$, $\gamma=0.1$, and $\mu=10^{-3}$ in the top row, $\mu=10^{-1}$ in the bottom row.
    }
    \label{fig6}
\end{figure}

Finally, we examine the spatial distribution of strategies on the lattice network, revealing the cooperation alliance, the underlying pathway for cooperation. Figure \ref{fig6} displays snapshots of the population evolution. Unlike the disordered mix seen at high exploration rates (the bottom row of Figure \ref{fig6}), the optimal rate enables the formation of cooperative alliances (the top row of Figure \ref{fig6}). Specifically, conditional cooperators form the protective outer shell of clusters, directly confronting defectors. Inside these protected enclaves, unconditional cooperators and specious cooperators thrive. This spatial organisation creates a shield, where the deliberative strategy (ACD) bears the cognitive cost to filter out defectors, thereby constructing a niche that supports a diversity of intuitive and deliberative cooperative strategies, until the internal exploitation by NDC triggers a local collapse and resets the cycle. It is this structural self-organisation that underpins the high cooperation levels and the cyclic dynamics observed in the time series.

\subsection{Exploration is not the panacea}
\begin{figure}[tb]
    \centering
    \includegraphics[width=\linewidth]{ 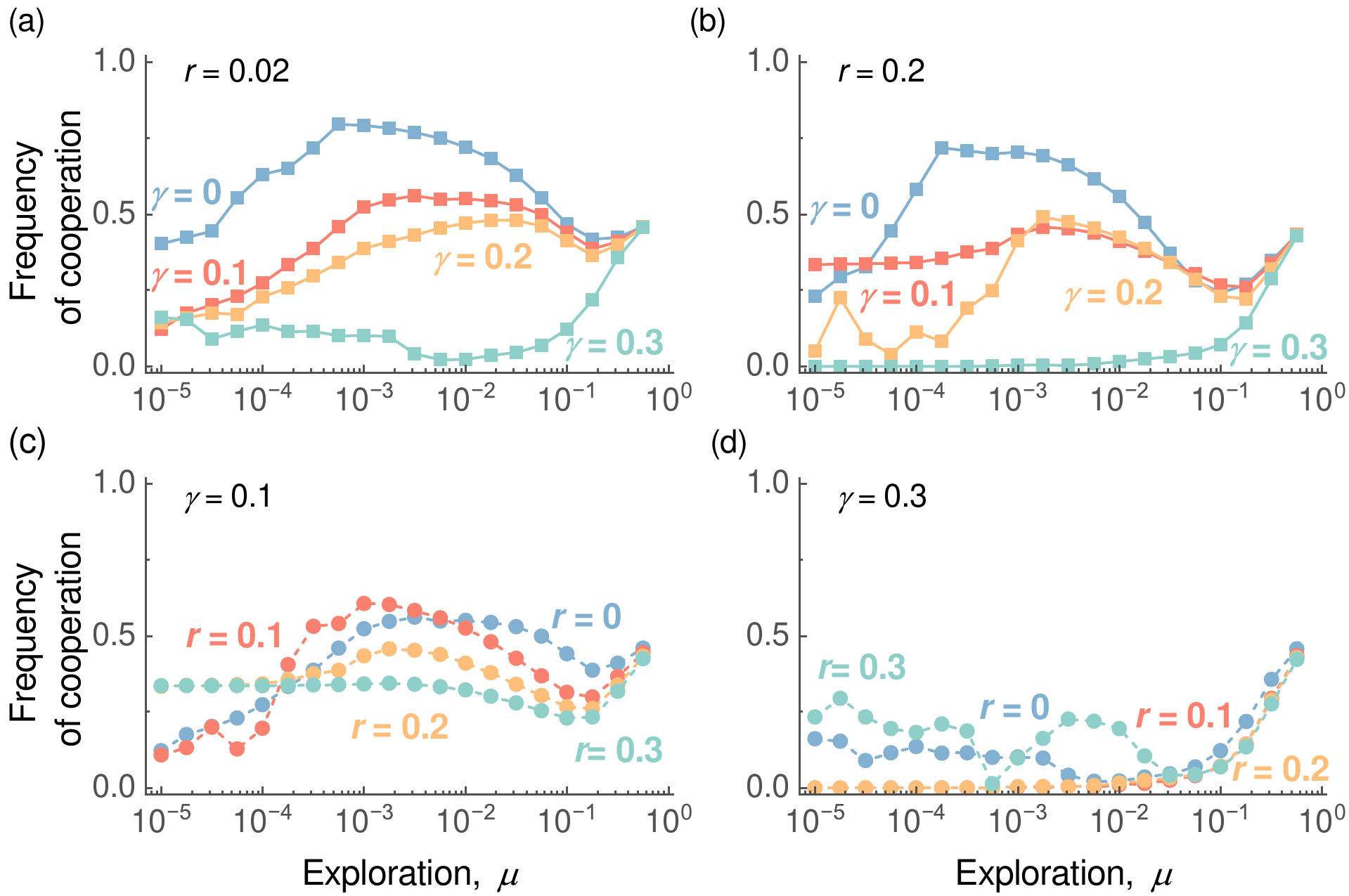}
    \caption{\textbf{Exploration can boost cooperative survival, but it is not a panacea. }
    Panels show the frequency of cooperation as a function of exploration on lattice network. 
    Parameters are set as: $r=0.02$ in (a), $r=0.2$ in (b), $\gamma=0.1$ in (c), and $\gamma=0.3$ in (d).
    }
    \label{fig3}
\end{figure}
While exploration is generally beneficial, its power to promote cooperation is conditional, failing to generate an optimal rate under strong dilemma strength and high reasoning cost.
Take lattice population as an example, first, when the reasoning cost is high (Figure \ref{fig3}(a-b)(d)), the benefit of exploration is suppressed. This high cost makes deliberative strategies using conditional logic extinct, leaving only costless unconditional strategies (C and D). Therefore, the system reverts to a conventional prisoner's dilemma, which exploration cannot rescue. Second, when the dilemma strength is strong (Figure \ref{fig3}(c)), the influence of exploration is also suppressed since the incentive to defect is too strong. The high dilemma strength prevents the formation of sustainable cooperative clusters, irrespective of the exploration rate. 
These results highlight that exploration can only unleash the cooperative potential of cheap talk when the underlying game parameters offer a chance for conditional strategies to survive; it cannot establish cooperation when either the dilemma strength or the cognitive cost is overwhelming.

\section{Discussion}
\label{Discussion}

Our findings align with a growing body of theoretical literature highlighting the constructive role of exploration in evolutionary games. Generally, exploration was often excluded for clear phase transitions or to simplify evolutionary dynamics. However, recent studies on direct reciprocity have suggested that exploration can actively enhance cooperation by preventing the system from freezing in defect-dominated absorbing states~\cite{tkadlec2023mutation}. Similarly, research on spatial evolutionary games has shown that exploration serves as a vital mechanism to mitigate finite-size effects, improving the reliability of agent-based models~\cite{shen2025mutation}. Our work contributes to this theoretical narrative by identifying a specific optimal exploration window within the context of cheap talk. Unlike pure action games, where exploration primarily refreshes the strategy pool, in our communication-based model, optimal exploration acts as a precise regulator: it provides enough volatility to destabilise rigid defection while remaining low enough to preserve the integrity of complex, conditional signalling strategies. This reinforces the theoretical view that strategic noise is not merely a disturbance to be minimised, but a fundamental parameter that drives system optimisation.

Beyond the primary focus on strategic exploration, our supplementary analysis reveals a counterintuitive phenomenon regarding the reasoning cost. While increasing the dilemma strength monotonically suppresses cooperation 
(Figures S2, S5, S8)
, the relationship between reasoning cost and the success of intuitive cooperators (ACC) follows a distinct inverted U-shape trajectory 
(Figures S3, S6, S9)
. This implies that a strictly zero-cost environment is not necessarily ideal for the proliferation of simple, unconditional cooperation. Mechanistically, a moderate reasoning cost acts as an evolutionary filter: it is sufficiently high to impose a selection pressure against complex, parasitic strategies (e.g., those that signal but defect) that exploit the communication channel, yet remains low enough to sustain a critical mass of conditional cooperators (ACD) who shield the community. In the context of engineering applications, reasoning cost corresponds to the computational overhead (e.g., CPU cycles, energy consumption, or processing latency) required for complex decision-making algorithms. Consequently, our findings suggest a design principle for resource-constrained systems: artificially imposing a marginal computational tax on complex logic, rather than striving for zero-latency processing, could paradoxically stabilise cooperation by rendering exploitative strategies evolutionarily unaffordable. Future work should verify this ``optimal cost hypothesis" for insightful system design guidance.

From a practical perspective, the existence of an optimal exploration rate offers valuable insights for the design of robust multi-agent systems and decentralised protocols. In many engineered systems, such as autonomous swarm robotics or peer-to-peer networks, designers often strive for deterministic efficiency, attempting to eliminate errors or deviations~\cite{brambilla2013swarm,lua2005survey}. However, our results suggest that imposing strict determinism can be counterproductive, potentially leading to system-wide traps of non-cooperation. Instead, adopting a protocol that incorporates a moderate level of stochastic exploration, functionally similar to the exploration-exploitation trade-off in Reinforcement Learning (RL), can be beneficial~\cite{leibo2017multi}. Just as RL agents occasionally deviate from their current policy to discover better global solutions, our model suggests that ``engineered randomness" allows the population to escape suboptimal local optima. For instance, in the coordination of autonomous vehicles, strict adherence to deterministic optimisation might lead to traffic gridlocks; however, allowing agents to occasionally deviate from the standard strategy can effectively break such symmetries and restore cooperative flow~\cite{wu2017emergent}. 

While our framework offers several insights, it relies on several simplifying assumptions that open avenues for future research. First, our simulations were conducted on various static networks to combine the effects of network reciprocity. Since real-world social and biological networks often exhibit time-varying properties, future work should investigate how dynamical interaction topologies influence the optimal exploration principle~\cite{li2020evolution}. 
Furthermore, our study primarily focuses on decentralised networks with relatively homogeneous degree distributions. While we observed that exploration disrupts cooperation in scale-free networks by potentially destabilising cooperative hubs, a granular analysis of how exploration specifically affects hub-versus-leaf dynamics remains an open question for future research~\cite{santos2005scale}.
Second, we limited the communication channel to a binary signal. In reality, communication is often multi-dimensional; expanding the model to include a richer diversity of signals could reveal whether increased semantic complexity, combined with exploration, further enhances cooperative stability~\cite{santos2011co}. Finally, our model assumes a uniform, global exploration rate for all individuals. However, in natural systems, the tendency to explore is often an evolving trait itself rather than a fixed constant. Future work could explore co-evolutionary models where exploration rates are adaptive, akin to meta-learning parameters in reinforcement learning, to understand how systems might self-tune to the optimal exploration level~\cite{szolnoki2009selection}.

\section{Conclusion}
\label{Conclusion}
To investigate how cheap talk sustains cooperation under the influence of exploration, we developed an evolutionary game framework featuring a two-stage cheap talk game on a spatially structured population.
We first confirmed that even in the presence of exploration, cheap talk combined with network reciprocity remains a powerful pathway for sustaining cooperation. Crucially, we discovered the existence of the optimal exploration rate that maximises the cooperative outcome, which facilitates the evolutionary success of both intuitive and deliberative cooperative strategies. Mechanistically, the moderate exploration rate works by undermining the stability of defection, yet enables cooperative strategies to form compact, resilient spatial clusters that utilise network reciprocity to resist invasion. However, strategic exploration is not a panacea; the optimal rate collapses, and cooperation remains suppressed under strong dilemma strength or excessively high reasoning cost, where the power of cheap talk fundamentally fails.

\section{Article information}
\paragraph*{Acknowledgments}
We acknowledge the support provided by EPSRC (grant EP/Y00857X/1) to Z.S. and T.A.H.

\paragraph*{Author contributions.} 
Z.\, S. and C.\,S conceived research. Z.\, S. performed theoretical analysis and evolutionary simulations. All co-authors discussed the results and wrote the manuscript.

\paragraph*{Conflict of interest.} Authors declare no conflict of interest.

\bibliographystyle{unsrt}
\bibliography{mybib}

\end{document}